\documentclass[aps,prl,reprint]{revtex4-2}

\usepackage{graphicx}% Include figure files
\usepackage{dcolumn}% Align table columns on decimal point
\usepackage{bm}% bold math
\usepackage{amsmath}
\usepackage{xcolor}
\usepackage{ulem}

\begin{document}

\preprint{APS/123-QED}

\title{Effective water/water contact angle at the base of an impinging jet}
\author{Théophile Gaichies}
 \author{Anniina Salonen}
\author{Emmanuelle Rio}%
 \email{emmanuelle.rio@universite-paris-saclay.fr}
\affiliation{%
Université Paris-Saclay, CNRS, Laboratoire de Physique des Solides, 91405, Orsay, France.
}%
\author{Arnaud Antkowiak}
\affiliation{Sorbonne Université, CNRS, Institut Jean le Rond $\partial$’Alembert, F-75005 Paris, France
}%

\date{\today}% It is always \today, today,
             %  but any date may be explicitly specified

\begin{abstract}

The base of a jet impinging on an ultrapure water bath is studied experimentally. At the impact point, a train of capillary waves develops along the jet. By performing Particle Tracking Velocity measurements, we show that there is a boundary layer separation between the jet and the meniscus. We thus describe the shape of this meniscus with a hydrostatic model. A striking observation is the existence of an effective non-zero water/water contact angle between the jet and the meniscus. The rationalization of this finite contact angle requires a full description of the shape of the interface. By doing an analytical matching between the meniscus and the jet, we show that the capillary waves can be considered as reflected waves present to ensure pressure continuity. It is finally shown that the value of the apparent contact angle is fixed by energy minimization, with an excellent agreement between prediction and experiment for small jets.

\end{abstract}

%\keywords{Suggested keywords}%Use showkeys class option if keyword
                              %display desired
\maketitle

%\tableofcontents

% %=================================
% % INTRODUCTION
% %=================================

Liquid-air interfaces are geometrical objects with an intimate link between their shape and their properties. 
They give rise to purely geometrical questions such as surface minimization problems \cite{courant1938existence,plateau1873statique}. 
A physical illustration is foams, whose mechanical stability is linked to the local bubble packing \cite{heitkam2012packing} and whose coarsening is fixed by topology \cite{von1952metal,hilgenfeldt2001accurate}.
Interfacial geometry can also play a critical role in dynamical situations. 
For example, the shape of the cusp at the rear of a sliding drop or an entrained film fixes the wetting transition \cite{le2005shape,snoeijer2006avoided}. 
More complex interfacial topologies exist in multiphase flows, such as breaking waves \cite{kiger2012air} or  aeration through  weirs or cascades \cite{ervine1998air}.
A paradigm for these flows is the plunging jet, which can entrain air when it impacts a bath \cite{kiger2012air}. 
Its geometrical features, such as the shape of the cusp in the viscous case \cite{eggers2001air,lorenceau2004air}, or surface disturbances when inviscid \cite{zhu2000mechanism}, control the whole behavior of the system.

\begin{figure}[h!]
    \centering
    \includegraphics[width=0.46\textwidth]{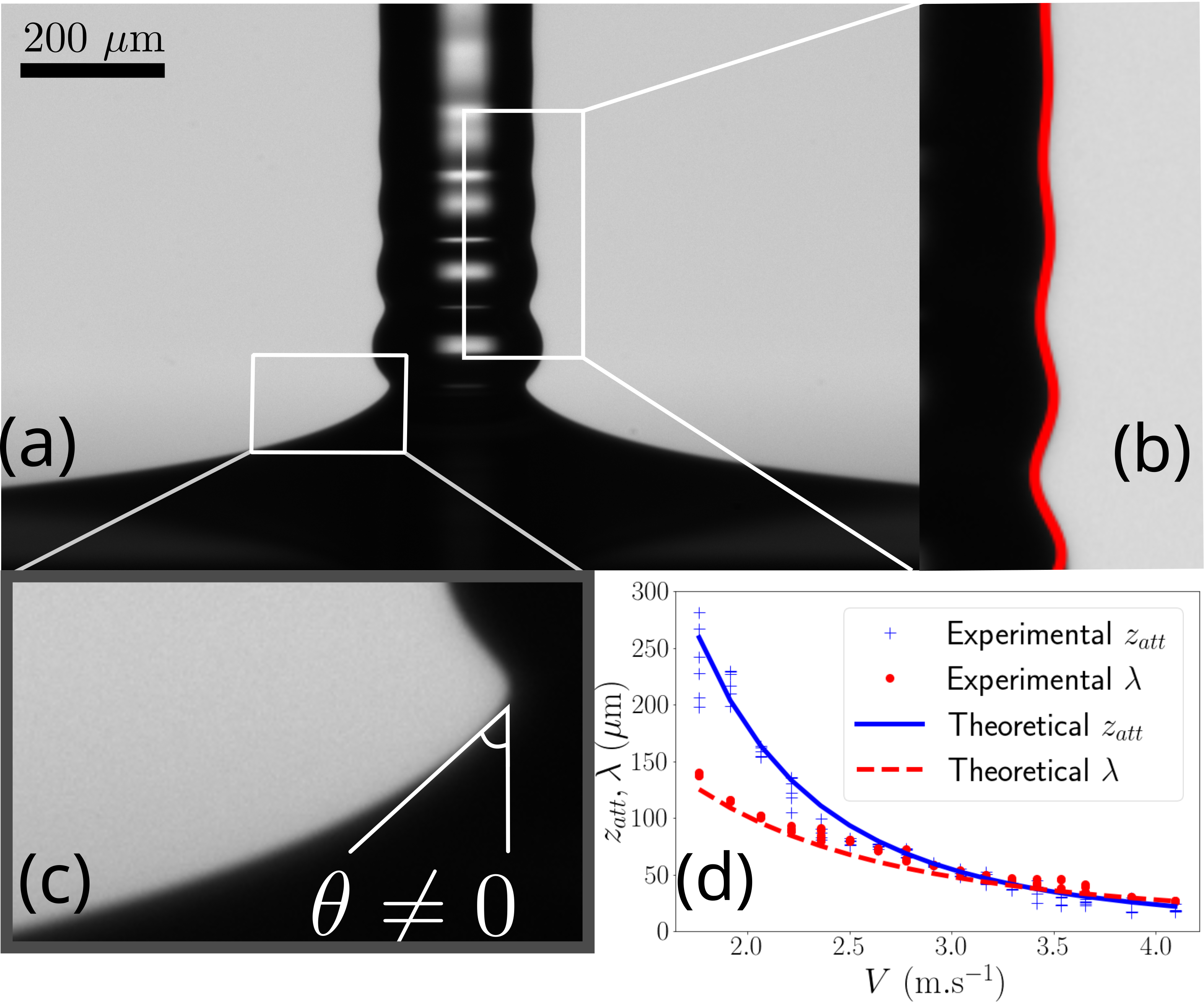}
    \caption{(a) Experimental image of a water jet of radius $R=100\ \mu$m and speed $V = 2.1$ $\mathrm{m.s^{-1}}$ impacting a bath of water. (b) Zoom on the capillary waves developing at the jet surface fitted by a function of the form $r(z)=R + a\ \cos(kz+\varphi) \exp(-\frac{z}{z_{\mathrm{att}}}) $. (c) Zoom on the connection between the meniscus and the jet, where the non-zero contact angle appears. (d) Comparison between the values of $z_{\mathrm{att}}$ and $\lambda$ extracted from the experimental fit and the ones calculated with Eq. 6 in ref.\cite{patrascu} for a jet of radius $R=160\ \mu$m.}
    \label{fig:intro}
\end{figure}

In this Letter, we focus on the base of water jets impacting a quiescent water bath. We report on an intriguing observation:  the apparition of an apparent non-zero contact angle at the water/water junction, see Fig.~\ref{fig:intro}(c).
We rationalize the presence of this contact angle and provide a description of the whole shape of this interface.

When a water jet plunges in a water bath, a train of stationary capillary waves will be created at its base, and decay upstream (Figure \ref{fig:intro}(a) and (b)) \cite{sklavenites1997wave}.
Their wavelength can be obtained by equating the phase velocity of the waves with the speed of the jet \citep{Lighthill1978}. 
The damping coefficient has
been modeled and linked to the properties of the fluid \cite{patrascu,awati1996stationary}. 
However, less attention has been given to the meniscus at the connection between the jet and the bath, where it is clear that there exists an apparent contact angle. This is surprising as the only phases at play are water and air, hence no contact angle is expected. 

To study this contact angle, we produce jets of ultrapure water (resistivity larger than 18.2 M$\Omega$.cm) with radii $R$ ranging from 50 to 400 $\mu$m, using a syringe pump (Harvard Apparatus PHD ULTRA) connected to a nozzle when $R < 100\ \mu$m and a pressurized reservoir otherwise. As shown in \cite{hancock2002fluid}, a small amount of surfactants in the recipient bath modifies significantly the wave pattern, so great care was taken to avoid contamination of the ultrapure water in our experiments.
The pressure in the reservoir is imposed by a pressure controller (Elveflow OB1 Mk4) connected to a network of compressed air. 
These setups can create laminar jets with speeds between $1.9$ and $8\ \mathrm{m.s^{-1}}$ for $R < 100\ \mu m$ and between $1.4$ and $5\ \mathrm{m.s^{-1}}$ for bigger ones. 
The base of the jet is imaged using telecentric lenses (SilverTl 2x or 4x Edmund Optics) to obtain sharp interfaces and a digital camera (Basler a2A1920-160umPRO). The images were typically taken with an exposure time of 700 $\mu$s. 

\begin{figure}[h!]
    \centering
    \includegraphics[width=0.46\textwidth]{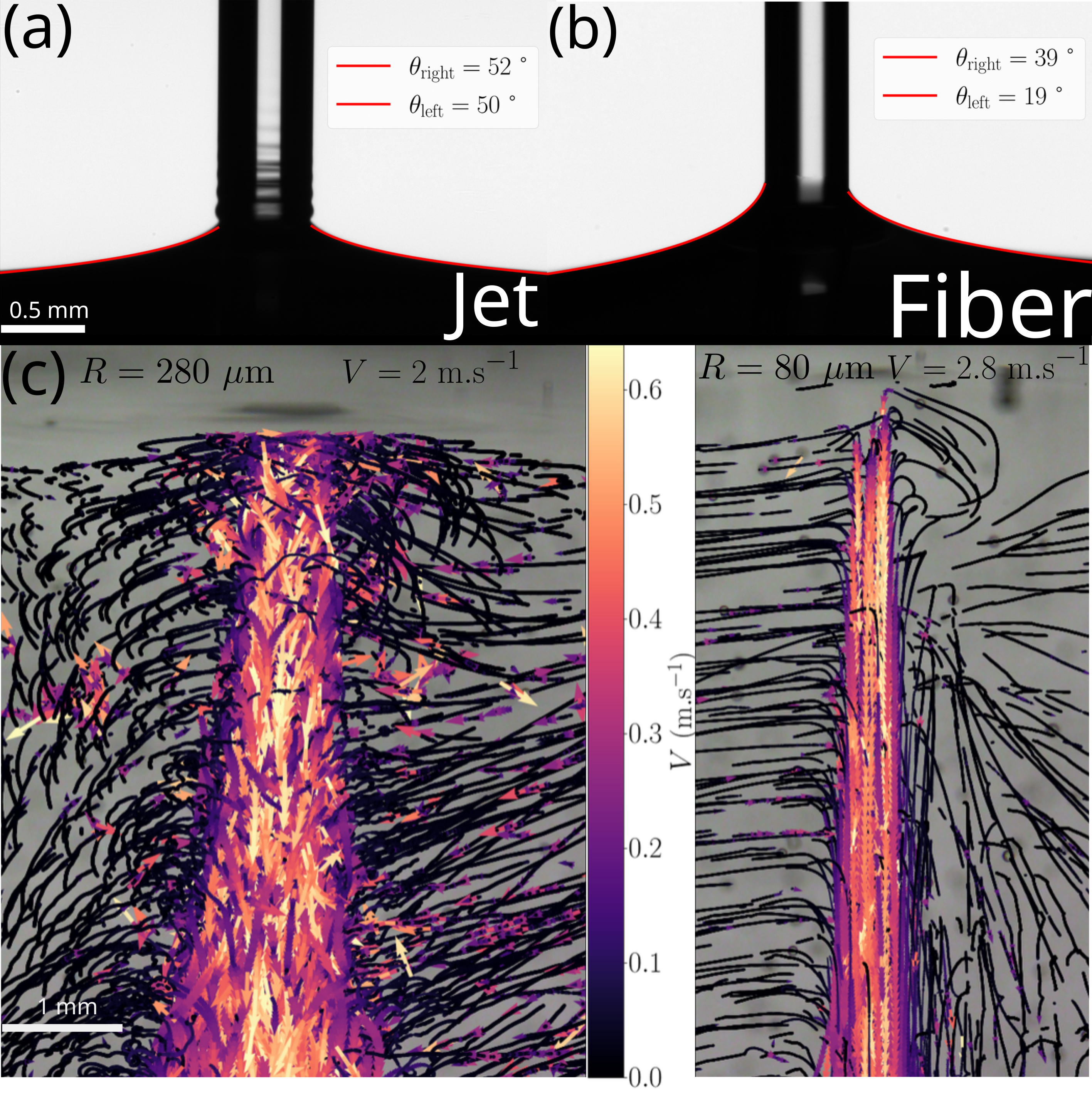}
    \caption{(a) Fit of the meniscus, on the right and on the left, by equation (1) for a jet of radius $R=280\ \mu$m and speed $V = 2.65 \ \mathrm{m.s^{-1}}$ and for (b) a glass fiber of $R=250\ \mu$m.  (c) Results of the PTV experiments under the water surface for a jet of  radius $R=280\ \mu$m and speed $V = 2 \ \mathrm{m.s^{-1}}$ and a jet of  radius $R=80\ \mu$m and speed $V = 2.8 \ \mathrm{m.s^{-1}}$}
    \label{fig:static}
\end{figure}

%=================================
% Caractérisation de l'angle de contact + envie de matcher 
%=================================
Measuring the observed contact angle is not straightforward, as it is not possible to place the contact line between the jet and the meniscus in a non-arbitrary manner.
To overcome this problem, we propose to measure the contact angle by fitting the shape of the interface.
Despite the jet having a speed of a few meters per second, the meniscus at the base of the jet looks very similar to the one climbing a static fiber plunged in a liquid bath (visual comparison in Fig.\ref{fig:static}(a) and (b)), as was previously noticed by Patrascu $et\ al.$ \cite{patrascu}. 
We performed PTV measurements below the interface, using small silica particles (of radius between 30 and 80 $\mu$m), and a fast camera (Photron FASTCAM NOVA S9) recording 6000 images per second. These images were then analysed with Python.
The results can be seen in Fig.\ref{fig:static}(c) and (d) for two different jet radii and velocities. 
The measurements confirm that there is a boundary layer separation between the dynamic jet and the static meniscus, with a typical velocity ratio of 20 between the velocity in the dynamic jet and in the still zone of the meniscus. 
This justifies the use of hydrostatic equations to describe the shape of the meniscus. 

The shape of the static meniscus is fixed by a balance between the hydrostatic pressure $\rho g z$, with $\rho$ the liquid density, $g$ the acceleration of gravity and $z$ the surface elevation and the Laplace pressure $\gamma \mathcal{C}$, with $\gamma$ the liquid/air surface tension and $\mathcal{C}$ the curvature of the interface. 
The typical length scale of the meniscus is thus the capillary length $\ell_c = \sqrt{\frac{\gamma}{\rho g}}$. 
An equation for $z(r)$, with $r$ the coordinate in the radial direction, in the non-linear part of the meniscus attached to a fiber of radius $r_0$ with a non-zero contact angle has been proposed by James \cite{james}. 
The linear small slope solution is $z(r)=AK_0(\frac{r}{l_c})$, where $A$ is a constant and $K_0$ a modified Bessel function of the second kind of order zero. 
By matching it with the equation of a catenoid satisfying the boundary condition $z'(r_0/r_0) = -\mathrm{cotan}(\theta)$ where $\theta$ is the contact angle, he obtains, in the vicinity of the fiber
\begin{equation}
\begin{split}
    z(r)=&r_0 \cos(\theta)\left[-\ln(\epsilon) + \ln(4) -\Gamma \right.\\
    &\left.-\ln\left(\frac{r}{r_0}+\left(\left(\frac{r}{r_0}\right)^2 - \cos^2\left(\theta\right)\right)^{\frac{1}{2}}\right)\right],
\end{split}
\end{equation}
where $\epsilon = \sqrt{\frac{\rho g R^2}{\gamma}}$ is the square root of the Bond number and $\Gamma$ is Euler's constant. 
We use James' equation to fit the meniscii in Fig. \ref{fig:static}(a) and (b).
The contact angle is used as a free parameter and can be different on the right and on the left.
This leads to slightly different values of the contact angle on the jet, which will be taken into account while evaluating the error bars. 
On the fiber, the difference is larger, probably due to impurities at its surface.
The agreement between the fit and the experimental shape confirms that this region can be considered as static, and that there exists a non-zero contact angle between the meniscus and the jet. 
Replacing directly the fiber radius $r_0$ in James' equation with the jet radius $R$ proves unsatisfactory, for the actual radius is altered with the capillary wavefield.
We thus need to write proper matching conditions between the capillary wavefield and the static meniscus.

In the following, we propose a matching between James' description of the meniscus and the jet decorated with capillary waves.
The parameters needed to describe the meniscus using James' equation are $r_0$ and $\theta$.
The capillary wavefield is approximated by an attenuated sinusoidal wave $r(z)=R + a\ 
\cos(kz+\varphi) \exp(-\frac{z}{z_{\mathrm{att}}}) $, with $k =\frac{2\pi}{\lambda}$ the wavenumber, $\lambda$ the wavelength, $a$ the amplitude, $\varphi$ a phase shift, and $z_{\mathrm{att}}$, the attenuation length. 
$k$ and $z_{\mathrm{att}}$ are predicted by the model of Patrascu (equation (6) in reference \cite{patrascu}). 
It agrees very well with the fitted values of $z_{\textrm{att}}$ and $\lambda$, as can be seen in Fig. \ref{fig:intro}d. 
We then have five unknowns $\theta,\ r_0,\ z_0,\ \varphi,\mathrm{and}\ a$. 
The first matching condition is the radial matching position, which gives $r_0=r(z_0) $. The height matching position $z_0$ is given by James' equation $z_0=z(r_0)$. 
The slope matching yields $\frac{\partial r}{\partial z}=-\tan(\theta)$. 
The last condition is the matching between the jet curvature $$\kappa_{\textrm{jet}}=\frac{1+r'(z)^2 -r(z)r''(z)}{r(z)[1+r'(z)^2]^{3/2}}$$ and the meniscus curvature $\kappa_{\textrm{m}}=0$, as the meniscus is a catenoid. 
Practically we used a linearized version of $\kappa_\text{jet}$ as full nonlinear computations did not show significant differences.

\begin{figure}[h!]
    \centering
    \includegraphics[width=0.46\textwidth]{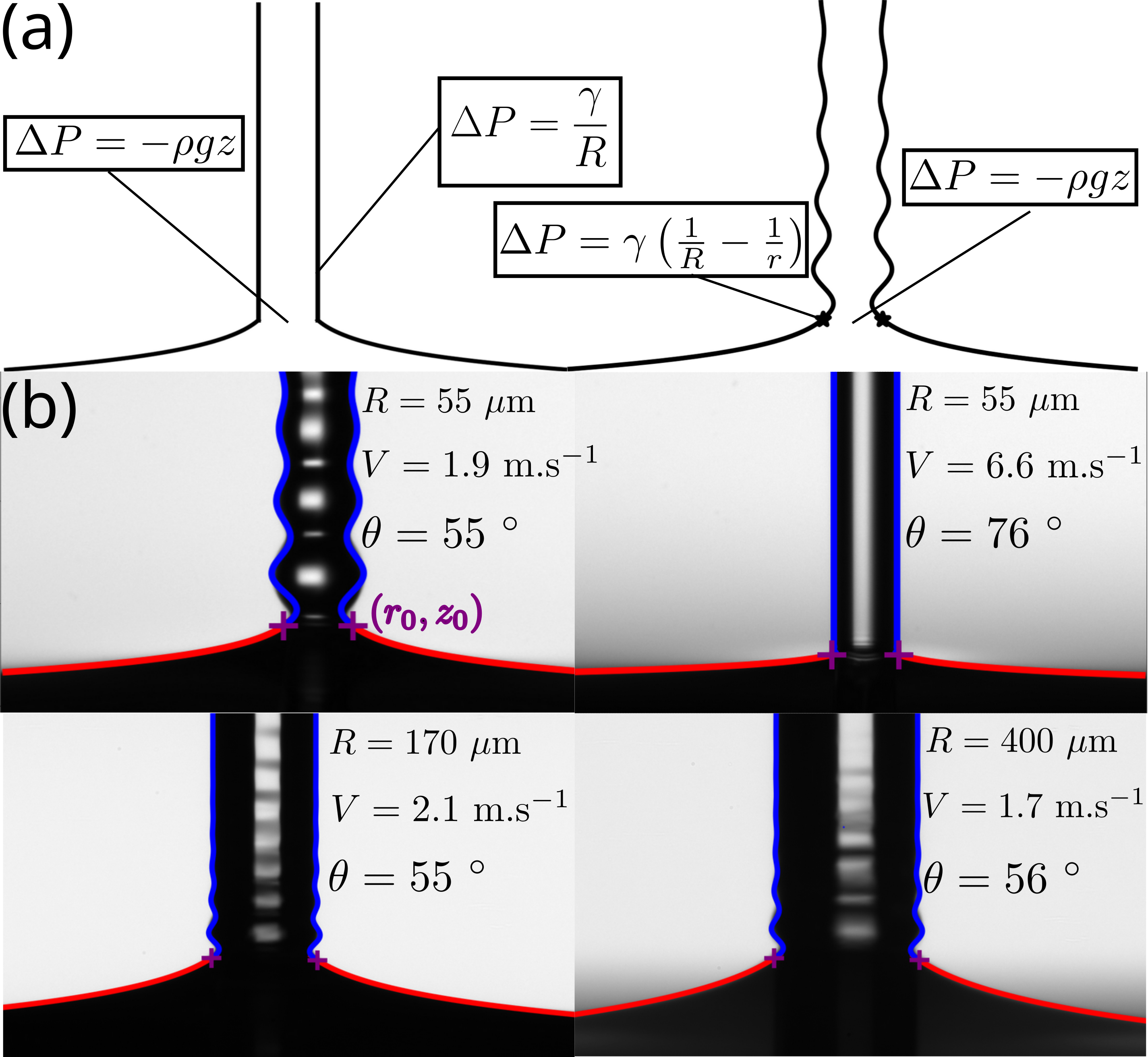}
    \caption{(a) Sketch of the jet impacting a bath. In the first virtual case, a straight jet connects to the meniscus, and exposes the impossibility of a pressure continuity between the two regions. In the second case, the capillary waves allow for pressure continuity. (b) Fitted interfaces of jets of various radii and speeds. The red line is the fitted meniscus, the blue line the wavefield with the calculated amplitude, and the violet cross the calculated matching point $(r_0,z_0)$.}
    \label{fig:fit_complet}
\end{figure}

This curvature matching is in fact a pressure matching between the jet, which has an overpressure $\Delta P=\frac{\gamma}{R}$, and the meniscus, which has an underpressure $\Delta P=-\rho g z$. 
Capillary wave emission stems from this pressure mismatch. The reflected waves lower the curvature of the impinged jet base to accommodate with the meniscus' pressure, as sketched in Fig.\ref{fig:fit_complet}a.
\ref{fig:fit_complet}a. 
This is similar to the stationary wave rippling the air cavity entrained behind an object dropped in a liquid\cite{belmontekeller2007cavity}.
In this case, a reflected stationary wave is needed so that the cavity is attached to the side of the moving object. 
These situations are reminiscent of classical problems in wave physics, where boundary conditions are impossible to meet without a reflected wave \citep{Billingham2000}.

With five unknowns but only four matching conditions, the problem is under-constrained as stated. However, if we set a $\theta$, we can get all the other values thanks to the following system: 
\begin{equation}
\begin{split}
     & r_0=\frac{R\left(R^2 + z_{\mathrm{att}}^2 (k^2R^2-2)+2z_{\mathrm{att}}R\tan(\theta)\right)}{R^2 + z_{\mathrm{att}}^2 (k^2R^2-1)}, \\
    & z_0=r_0\cos(\theta)\left(\log(\frac{4R}{r_0\epsilon(1+\sin(\theta))})-\Gamma\right), \\
    & \varphi = -kz_0 + \arctan\left(\frac{r_0 - R - z_{\mathrm{att}}\tan(\theta)}{kz_{\mathrm{att}}(R-r_0)}\right), \\
    & a = \frac{(r_0-R)\exp(z_0/z_{\mathrm{att}})}{\cos(kz_0+\varphi)},
\end{split}
\label{eq:matching}
\end{equation}
where $\varphi$ can be shifted by $\pi$ to get a positive amplitude.  
To test that our equations of matching indeed set the amplitude of the capillary waves, we follow this fitting procedure: for a jet of given $R$ and $V$, we first get $k$ and $z_{\mathrm{att}}$ by fitting the wave pattern at the base of the jet, or from Patrascu equation (that only depends on $R$ and $V$) when the waves become barely visible on our images. 
This fixes $r_0(\theta)$ through Eq. \ref{eq:matching}.
We then obtain $\theta$ by fitting the meniscus with  James' equation. 

When $\theta$ is fitted, we  inject it in system \ref{eq:matching} and deduce the values of $z_0,\ \varphi$ and $a$. 
In Fig.\ref{fig:fit_complet}b, the resulting wavefield is plotted on the image, together with James' equation for the fitted $\theta$. 
We can see that, for various radii and jet velocities, the obtained amplitude agrees very well with the experimental one. 
This confirms that the pressure matching between the meniscus and the jet sets the amplitude of the capillary wavefield. 

\begin{figure}[h!]
    \centering
    \includegraphics[width=0.46\textwidth]{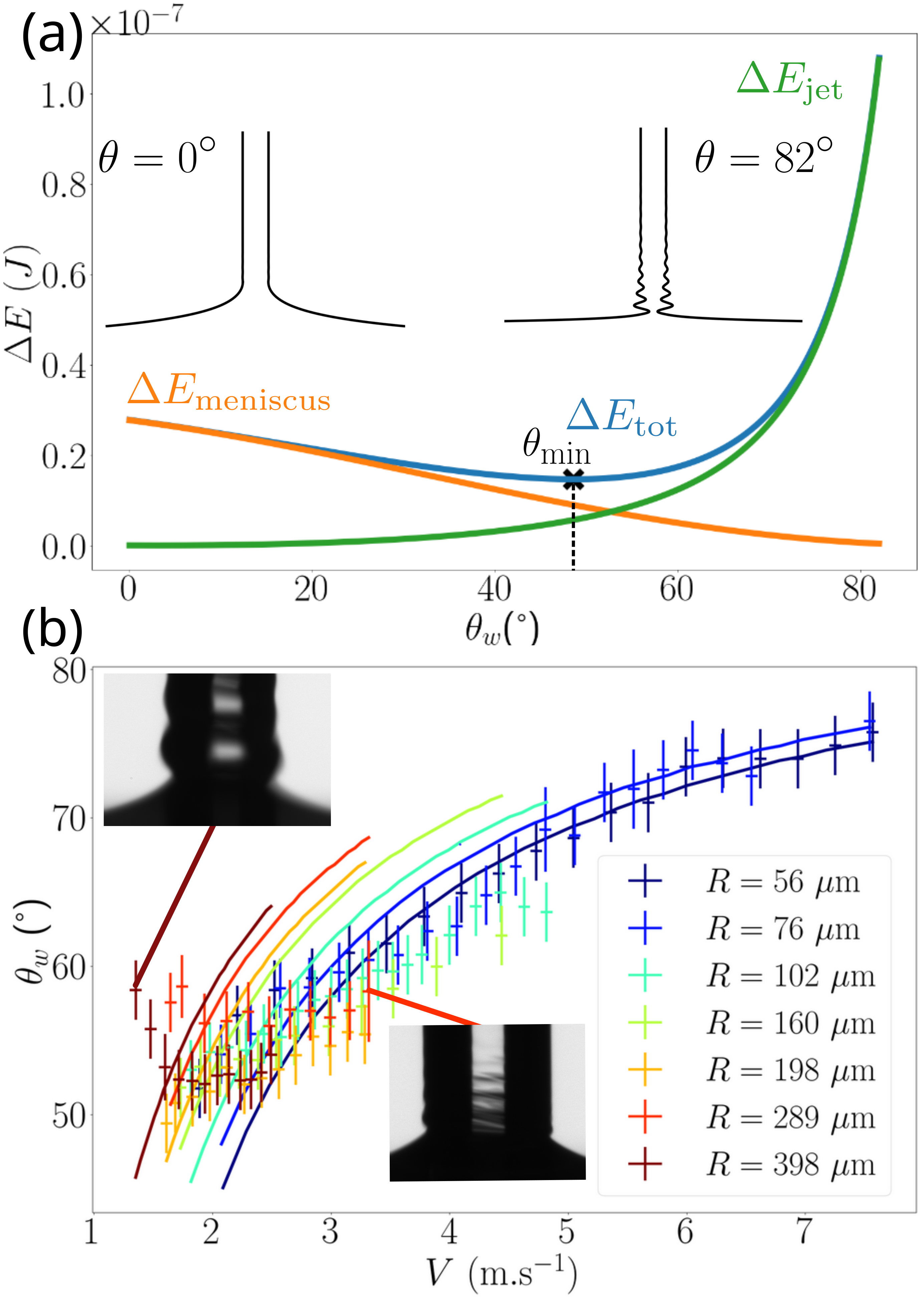}
    \caption{(a) Energy excess of the meniscus ($\Delta E_{\mathrm{jet}}$), of the jet ($\Delta E_{\mathrm{meniscus}}$) and their sum ($\Delta E_{\mathrm{tot}}$) plotted against the speed for a jet of radius $R=190\ \mu$m. The inserted sketches show the difference between the shape of the interface for a low and high contact angle. 
    (b) Apparent contact angle fitted and predicted by our model for jets of various radii, plotted against the speed of the jet. 
    The two experimental images show two reasons for the deviation from our model, the first one being instationarity (at the top) and the second one (at the bottom) the loss of axisymmetry, even when the stationarity is restored.}
    \label{fig:theo_vs_exp}
\end{figure}

To get a complete model of the interface, we miss a prediction for $\theta$, as all $\theta$ are compatible with our equations of matching. 
However, the total energy of the system depends on $\theta$. 
As can be seen in Fig.\ref{fig:theo_vs_exp}a, when $\theta$ is small, the meniscus joins the jet at a higher altitude, resulting in an excess surface compared to a flat meniscus. 
When $\theta$ is large, the amplitude of the capillary waves is larger, so the excess of surface compared to an unperturbed jet is higher. 
The kinetic energy excess caused by the capillary waves is equal to the surface energy excess according to Chandrasekhar (equation (154) and (57) of chapter XII in \cite{chandrasekhar2013hydrodynamic}). 
It is then possible to calculate the difference in energy $\Delta E$ with a virtual cylindrical waveless jet and flat meniscus as a function of $\theta$: 
\begin{equation}
\begin{split}
    \Delta E_{\mathrm{tot}}(\theta)=\ 2\pi\gamma & \left( 2\int_{z_0(\theta)}^{\infty}[R(z,\theta)\sqrt{1+\frac{\partial R(z,\theta)}{\partial z}^2}-R]dz \right. \\
    & \left. +\int_{r_0(\theta)}^{\infty}r(\sqrt{1+\frac{\partial z(r,\theta)}{\partial r}^2}-1)dr  \right) \\
    & + \rho g\int_{r_0(\theta)}^{\infty}\pi r z(r,\theta)^2dr,
\end{split}
\end{equation}
where the last term accounts for the gravitational potential energy of the meniscus \footnote{The meniscus energy term has here been computed using the uniformly valid composite solution of James, see \S 6 of \citet{james}}.
The sum of the jet and meniscus energies is plotted in Fig. \ref{fig:theo_vs_exp}a, revealing a value $\theta_{\mathrm{min}}$ which minimizes $\Delta E_{\mathrm{tot}}$.
This $\theta_{\mathrm{min}}$ depends on $V$ and $R$ as these variables set $k$ and $z_{\mathrm{att}}$, which affect $\Delta E_{\mathrm{jet}}$. 
In Fig.\ref{fig:theo_vs_exp}b, the prediction of $\theta_{\mathrm{min}}$ is plotted together with the experimental data as a function of $V$ for different radius. The uncertainty for the experimental value of $\theta_{\mathrm{w}}$ comes from the numerical uncertainty on the fit of the meniscus. 
For small jets ($R=56\ \mathrm{\mu m}$ and $R=76\ \mathrm{\mu m}$), the model agrees well with the experiments without any fitting parameters, especially when $V > 3 \ \mathrm{m.s^{-1}}$. 
For lower speeds, $a/R$ is not negligible and capillary waves are not well described by an attenuated sinusoid \cite{Vanden-Broeck1998} (see the first image in Fig.\ref{fig:fit_complet}b), so we expect a bigger difference with our model. For wider jets, non stationarities, and a certain degree of swirl \cite{patrascu} can appear (see Fig.\ref{fig:theo_vs_exp}b), which result in deviations from our model, as our assumptions of stationarity and axisymmetry are no longer fulfilled. 

In conclusion, the observation of an apparent contact angle when a jet impacts a bath has been rationalized. Capillary waves are due to a pressure matching between the jet and the meniscus, and energy minimization fixes the effective contact angle at least for small jets.
Our data are in excellent agreement with this prediction for tiny jets. 
Discrepancies appearing for wider jets underline the necessity to better understand the non-stationary and asymmetric regimes.

\bibliographystyle{unsrt}
\bibliography{references}

\end{document}